\documentclass[prb,twocolumn,amsmath,amssymb,superscriptaddress,citeautoscript,footinbib]{revtex4-1}
\usepackage[utf8]{inputenc}
\usepackage{color}
\usepackage{xcolor}
\usepackage{amsmath}
\usepackage{amssymb}
\usepackage{geometry}
\usepackage{indentfirst}
\usepackage{epsfig}
\usepackage{graphicx}
\usepackage{subfigure}
\usepackage{systeme}
\usepackage{verbatim}
\usepackage{bm}
\usepackage{graphicx}

\begin{document}
\title{Point defect evolution under irradiation: finite size effects and spatio-temporal correlations}

\author{Enrique Mart\'inez}
\email{enriquem@lanl.gov}
\affiliation{Theoretical Division, T-1, Los Alamos National Laboratory, Los Alamos, 87545 NM, USA}

\author{Fr\'ed\'eric Soisson}
\email{frederic.soisson@cea.fr}
\affiliation{DEN-Service de Recherches de M\'etallurgie Physique, CEA, Université Paris-Saclay, F-91191 Gif-sur-Yvette, France} 
\author{Maylise Nastar}
\email{maylise.nastar@cea.fr}
\affiliation{DEN-Service de Recherches de M\'etallurgie Physique, CEA, Université Paris-Saclay, F-91191 Gif-sur-Yvette, France}

\date{\today}


\begin{abstract}
The evolution of point defect concentrations under irradiation is controlled by their diffusion properties, and by their formation and elimination mechanisms. The latter include the mutual recombination of vacancies and interstitials, and the elimination of point defects at sinks. We show here that the modelling of this evolution by means of atomistic kinetic Monte Carlo (AKMC) simulations, necessarily using small system sizes, introduces strong space and time correlations between the vacancies and interstitials, which may strongly affect the recombination rate and the point defect concentrations. In such situations, standard rate theory models fail to predict the actual point defect concentrations. The effect is especially strong when the elimination of point defects occurs only by recombination, but can still be significant in the presence of sinks. We propose a new Correlated Pair Theory that fully takes into account the correlations between vacancy and interstitial pairs and predicts point defect concentrations in good agreement with AKMC simulations, even in very small systems. The Correlated Pair Theory can be used to modify the elimination rates in AKMC simulations to yield point defect concentrations as predicted by the standard rate theory, i.e. representative of large systems, even when using small simulation boxes.  
\end{abstract}

\maketitle

\section{Introduction}

In  materials under irradiation, the elastic collisions between the irradiation particles and the atoms of the material produce vacancies and self-interstitials. These point defects can migrate by thermally activated diffusion and eliminate by mutual recombination or annihilation at sinks such as dislocations, grain boundaries, free surfaces or point defect clusters. The balance between these competing mechanisms results in defect concentrations that may exceed the equilibrium values by orders of magnitude \cite{Sizmann1978,Was2007}.

Since the diffusion coefficients of substitutional atoms are proportional to point defect concentrations, a direct effect of irradiation is the acceleration of diffusive phase transformations, such as the precipitation of secondary phases or the ordering of intermetallic phases. In addition, fluxes of excess point defects towards sinks lead to other phenomena such as radiation-induced segregation, void swelling, irradiation creep and growth \cite{Was2007}. All these phenomena may have a significant (and usually deleterious) impact on the material properties. The reliable prediction of vacancy and interstitial concentrations is therefore a critical part of the modelling of irradiation effects. Since the pioneering work of Dienes and Damask \cite{Dienes1958}, and Lomer \cite{Lomer1959}, the most simple way to model the evolution of point defect concentrations is to used rate theories that account for the different formation and annihilation mechanisms (see Refs.~\onlinecite{Sizmann1978} and \onlinecite{Was2007} for detailed reviews). These models introduce averaged concentrations that do not fully take into account the time and space correlations between point defects.
On the other hand, atomistic models, such as molecular dynamics (MD) and atomistic kinetic Monte Carlo (AKMC) simulations take fully into account these correlations. Since they are usually limited to relatively small systems (say $N_s \sim 10^5$ to $10^7$ atoms) and since the point defect concentrations remain small at a given time (even when we take into account the increase due to irradiation), the average number of point defects in the system (vacancies $N_v$ and self-interstitials $N_i$) can be close to or even below one. In such case,
finite size effects on the formation and annihilation reactions of point defects become critical.

One can consider this situation from two points of view. If one deals with actual small systems (defined as $N_v,N_i \ll 1$), one can use atomistic simulations to test the rate theory, and  suggest a way to include an explicit treatment of the spatio-temporal correlations between point defects. On the contrary, if one wants to model big systems (where $N_v,N_i \gg 1$) with small simulation boxes (probably the most frequent case), one may try to modify the atomic diffusion model in order to get the spatio-temporal correlations -- and therefore point defect concentrations -- that are more representative of the real situation. The objective of this paper is to deal with these issues. We start with a quick summary of the standard rate theory (SRT), and of its predictions concerning the steady-state concentrations of point defects (section \ref{Standard Rate theory}). In section \ref{Monte Carlo Simulations and finite size effects}, we present the AKMC simulations of pure body centered cubic (bcc) iron under irradiation, first in a system without sinks (where the elimination of point defects occurs only by recombination). In this case, one easily shows that SRT leads to an underestimation of the recombination rate and to an overestimation of the point defect concentrations that can reach orders of magnitudes for small system sizes. When sinks are present, we show that the discrepancy is less spectacular, but can be significant. In section \ref{toto}, we propose a revised rate theory (referred to as Correlated Pair Theory, or CPT) that gives a better description of space and time correlations and yields point defect concentrations in very good agreement with AKMC results. In section \ref{Modified AKMC} we discuss how to modify the AKMC algorithm in order to reproduce the correlations and point defect concentrations of a big system, in a small simulation box. Finally, we discuss the relevance of these issues depending on the properties of the materials and the irradiation conditions.

\section{Standard Rate theory} \label{Standard Rate theory}

In SRT models \cite{Sizmann1978}, the evolution of the average vacancy ($v$) and self-interstitial ($i$) concentrations under irradiation, $X_{v}$ and $X_{i}$, are  given by chemical rate equations:

\begin{equation}
\begin{split} 
\frac{dX_{v}}{dt} & =G'-K_{iv} X_{i} X_{v}-\sum_s k_{sv}^2 D_v X_{v}\\
\frac{dX_{i}}{dt} & =G'-K_{iv} X_{v} X_{i}-\sum_s k_{si}^2 D_i X_{i}
\end{split}
\label{SRT}
\end{equation}

The first term in the RHS of Eqs.~(\ref{SRT}), $G'$, is an effective production rate. When the kinetic energy $E_0$ transferred to the primary knock-on atom (PKA) exceeds a threshold energy $E_d$, the number of Frenkel pairs initially produced by elastic collisions is usually estimated by the standard NRT model \cite{Was2007}: $n=(E_0-E_i)/(2E_d)$, where $E_i$ is the energy lost in inelastic interactions. The corresponding production rate $G$ gives the number of displacement per atom (NRT dpa) per second. 

During ion or neutron irradiations ($E_0 \gg E_d$), Frenkel pairs are created in localized displacement cascades and many close $i-v$ pairs will immediately recombine or form point defect clusters, leading to a lower effective production rate $G'=\xi G$. During electron irradiation, Frenkel pairs are created more homogeneously. Nevertheless, the initial distance between the two defects of a Frenkel pair is sufficiently small to produce close-pair recombination after a few point defect jumps. It can be shown that in this case $\xi=1-R_{rec}/R_p$, where $R_{rec}$ is the distance of recombination and $R_p$ the initial average $i-v$ distance \cite{Schroeder75}. 

This kind of close-pair correlations -- either in ion, electron or neutron irradiations --  are well-known physical correlations that do not depend on the size of the system. They lead to the correction of the production rate by a constant factor. They may also affect point defects clustering within the cascades and resistivity recovery experiments \cite{Ortiz2007,Terentyev2012}. It should be noticed that they are different from the correlations studied in the present study, which result from the finite size of the system used in the simulations.

The second term in Eqs.~(\ref{SRT}), $K_{iv} X_{i} X_{v}$, is the recombination rate, derived from Waite's theory of the kinetics of diffusion-limited reactions \cite{Waite1957}:

\begin{equation}
K_{iv}=4 \pi R_{rec} \frac{D_i+D_v}{V_{at}},
\label{Waite}
\end{equation}

\noindent where $D_i$ and $D_v$ are the diffusion coefficients of point defects and $V_{at}$ the atomic volume.

The last terms in Eqs.~(\ref{SRT}), $\sum_s k_{sv}^2 D_v X_{v}$ and $\sum_s k_{si}^2 D_v X_{v}$, are the rates of elimination at point defect sinks. Each kind of sink $s$ is characterized by its sink strength $k_{sv}^2$ and $k_{si}^2$ \cite{Nichols1978}. If one neglects the sink biases ($k_{sv}^2$=$k_{si}^2$), Eqs.~(\ref{SRT}) become

\begin{equation}
\begin{split} 
\frac{dX_{v}}{dt} & =G'-K_{iv}X_{v}X_{v}-K_v X_{v},\\
\frac{dX_{i}}{dt} & =G'-K_{iv}X_{v}X_{i}-K_i X_{i},
\end{split}
\label{SRT2}
\end{equation}

\noindent with $K_d = \sum_s k_{sd}^2D_d$. 

One easily shows that at steady-state, without point defect sinks ($K_d =0$), the point defect concentrations are

\begin{equation}\label{eq:RTNoS}
X_v^{st}=X_i^{st}=\sqrt{\frac{G'}{K_{iv}}}
\end{equation}

In the presence of point defect sinks \cite{Sizmann1978}, we have

\begin{equation}
X_v^{st}=-\frac{K_i}{2K_{iv}}+\left[\left(\frac{K_i}{2K_{iv}}\right)^2+\frac{G'K_i}{K_{iv}K_v}\right]^{1/2}
\label{SSC_V}
\end{equation}

\begin{equation}
X_i^{st}=-\frac{K_v}{2K_{iv}}+\left[\left(\frac{K_v}{2K_{iv}}\right)^2+\frac{G'K_v}{K_{iv}K_i}\right]^{1/2}.
\label{SSC_I}
\end{equation}

We deduce from the steady state point defect concentrations, the ratio of the total point defect recombination rate divided by the elimination rate of both vacancy and interstitial at sinks
\begin{equation}
\frac{n_{rec}^e}{n_{sink}^e} = \frac{K_{iv} X_i^{st} X_v^{st}}{K_v X_v^{st} + K_i X_i^{st}}.
\label{ratiocor1}
\end{equation}

Many theoretical studies have been devoted to the calculation of sink strengths, and specific expressions have been established for the annihilation of point defects at grain boundaries, dislocations, cavities, etc. \cite{Nichols1978}. 
Most of the recombination models focus on the intra-pair reactions. They consider the population of point defects as a sum of isolated Frenkel pairs, and deal with the bimolecular recombination reaction of an isolated pair \cite{KOILA19741259,KOFMAN1978217,Schroeder75}. Authors such as Kotomin and Kuzovkov, tackled the spatial correlations of a many-point particle density \cite{Kotomin_1992}. They rely on a continuous mean-field approach to investigate the spatial fluctuations of the particle population. Although they account for pair spatial correlations, their approach does not give access to the interplay between finite size effects and time correlations of the point defect distribution on the probability of recombination reactions.   
 
 The expression $-K_{iv}X_v^{st} X_i^{st}$ in Eqs.~(\ref{SRT}) relies on the assumption that the probability of finding a pair $i-v$ is proportional to the product of the averaged point defect concentrations (or atomic fractions). This is a mean-field approximation that neglects the time and spatial correlations between the two kinds of defect. It is only justified on a large and homogeneous system, for which $X_v=N_v/N_s$ and $X_i=N_i/N_s$ are defined over large numbers of vacancies $N_v$ and self-interstitials $N_i$ in $N_s$ sites. In atomistic simulations, if one deals with realistic point defect concentrations and limited system sizes, these conditions are not always met, as shown in the following sections.

\section{Monte Carlo Simulations and finite size effects} \label{Monte Carlo Simulations and finite size effects}

In this section, we compare the steady-state point defect concentrations predicted by the SRT with those measured in AKMC simulations, in a simple model of pure bcc iron under irradiation.

\subsection{Diffusion model and Monte Carlo simulations}

We rely on the AKMC simulations and the diffusion model developed for the study of segregation and precipitation in Fe-Cr bcc alloys under irradiation \cite{SENNINGER20161,Soisson2018}. We use Monte Carlo boxes from $N_s=2\times64^3$ to $N_s = 2\times256^3$ bcc sites and periodic boundary conditions. The various events are chosen using a residence time algorithm (RTA). Details can be found in Refs. \onlinecite{SENNINGER20161} and \onlinecite{Soisson2018}.

\subsubsection{Diffusion}

Vacancy diffusion occurs by jumps towards one of the eight nearest-neighbor sites (jump distance $\lambda=\sqrt{3}a/2$, where $a=0.287$~nm is the bcc-Fe lattice parameter). Self-interstitial atoms have a $\left<110\right>$-dumbbell configuration and can jump towards four of the eight nearest-neighbor sites with a $60^\circ$ rotation, according to the Johnson's mechanism \cite{Johnson1964}.  

Following transitions state theory, the $v$ jump frequency in pure Fe is: $\Gamma_v=\nu_0 \exp(S^m_v/k_B) \exp(-H^m_v/k_BT)$, with constant entropy and enthalpy of migration, $S^m_v=2.1 k_B$ and $H^m_v=0.690$~eV. Similarly for the $i$ jump frequency: $\Gamma_i=\nu_0 \exp(S^m_i/k_B) \exp(-H^m_i/k_BT)$ with $S^m_i=2.1 k_B$ and $H^m_v=0.343$~eV. The attempt jump frequency is $\nu_0 = 10^{13}\rm s^{-1}$ for both defects.

\subsubsection{Formation}

For the sake of simplicity, the mechanism of formation of vacancies and interstitials used in the AKMC simulations are representative of a damage created under electron irradiations. The Frenkel pairs are introduced individually within the system, with a frequency $G$ per bcc site (i.e. with a dose rate $G$ in $\rm dpa.s^{-1}$): one vacancy $v$ is created by removing a randomly chosen atom, and one interstitial $i$ is created at a distance $R_p$ in one of the eight $\left<111\right>$ directions, randomly chosen. To illustrate the effect of close pair recombinations, we have used two values for $R_p$, $R_p = 20\lambda$ and $R_p=7 \lambda$. 

 During ion or neutron irradiations, several Frenkel pairs are simultaneously created  within small displacement cascades, and small point defect clusters are also formed -- which later act as point defect sinks -- and are not considered in the present study.

\subsubsection{Recombination and Elimination at sinks}

After each jump, a point defect immediately recombines with an opposite defect if their distance is below the recombination distance $R_{rec}=3a$.

To model the point defect elimination at sinks, we use two approaches: 

- We introduce in the simulation box, a random distribution of point-like sinks, with an atomic fraction $C_s$. The corresponding sink strength can be evaluated by measuring the steady-state values $X_v^{st}$ and $X_i^{st}$ in AKMC simulations with various $C_s$ and no recombination. For $C_s > 1/N_s$, one finds that $k ^2_v=k^2_i=8C_s/(1.3a^2)$ \cite{Soisson2018}, independently of temperature, dose rates, or point defect properties.

- We do not introduce a static distribution of sinks, but an annihilation rate given by $K_d$ per defect for each type of defect that is sampled by the AKMC in the same way as hopping rates.\cite{marian_stochastic_2011}

In the simple problem considered here (point defect concentrations in pure metals), both methods give very similar results. The main advantage of the second method is that it does not introduce finite size effects on the elimination rate at sinks, even when $C_s<1/N_s$. Both methods can be combined to study complex phenomena such as radiation-induced segregation, requiring realistic geometries \cite{SENNINGER20161,Martinez2018}. 

\subsubsection{Point defect concentrations}

We measure the point defect concentrations by computing the average values of $N_v/N_s$ and $N_i/N_s$ over a sufficient number of Monte Carlo steps (MCS). We take advantage of the RTA algorithm by weighting every MC configuration by its residence time. These averaged concentrations can be directly compared with the solution of SRT, using Eqs.~\ref{SSC_V} and \ref{SSC_I} with $G'=(1-R_{rec}/R_p)G$, $D_v = a^2\Gamma_v$, $D_i = a^2\Gamma_i/2$, $V_{at}=a^3/2$ and $k^2_v=k^2_i=8C_s/(1.3a^2)$. 

\begin{figure}[htbp]
\centering
\includegraphics[width=1.0\columnwidth]{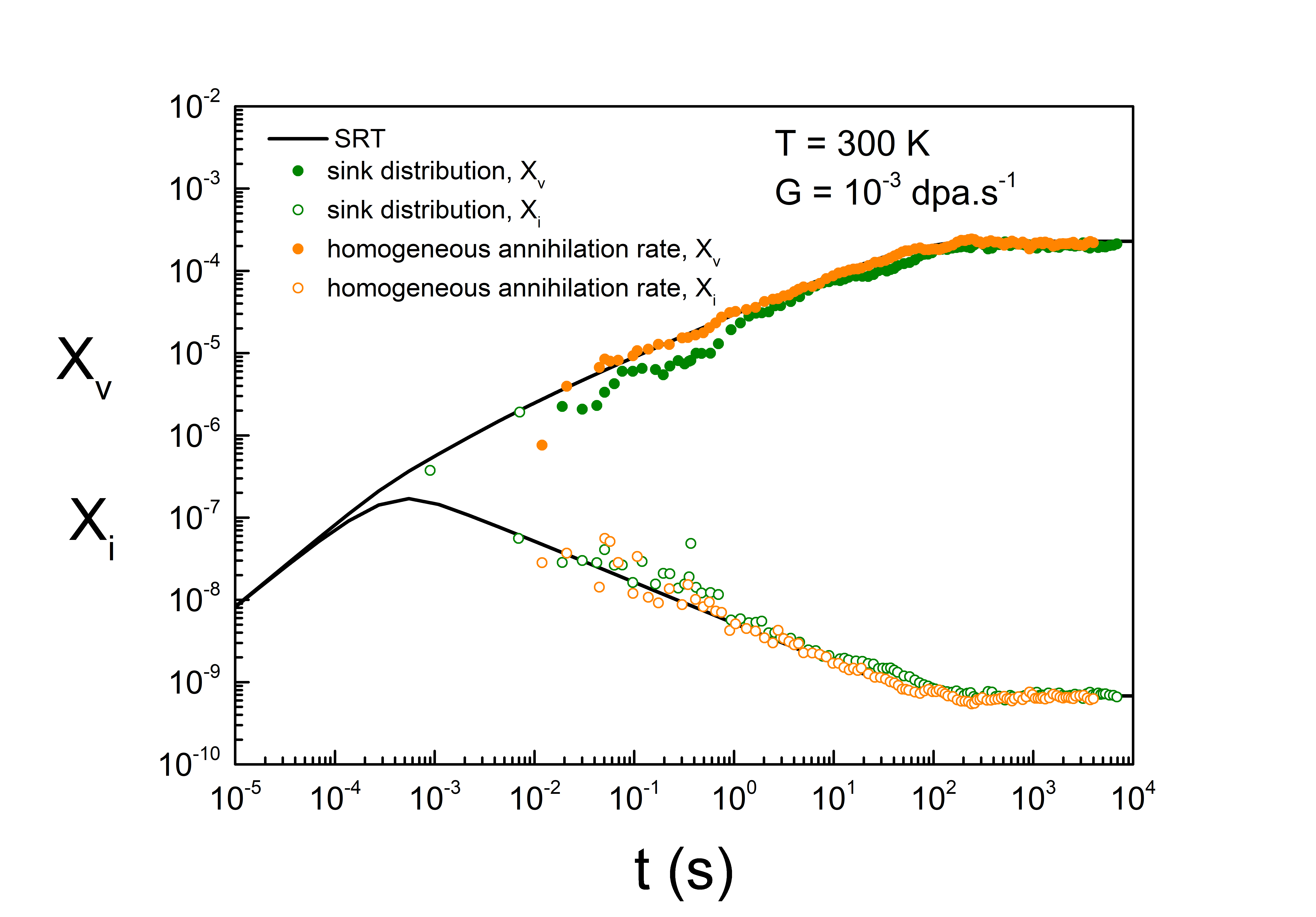}
\caption{Evolution of point defect concentrations at $T=300$~K and $G=10^{-3} \rm dpa.s^{-1}$, $k^2_ i=k^2_ v=5\times 10^{10}\rm cm^{-2}$. AKMC simulations with $R_p=20\lambda$ and $N_s=2\times64^3$, green dots: AKMC with a real distribution of sinks, orange dots: AKMC with an average elimination rate.}
\label{fig1}
\end{figure}

\begin{figure}[htbp]
\centering
\includegraphics[width=1.0\columnwidth]{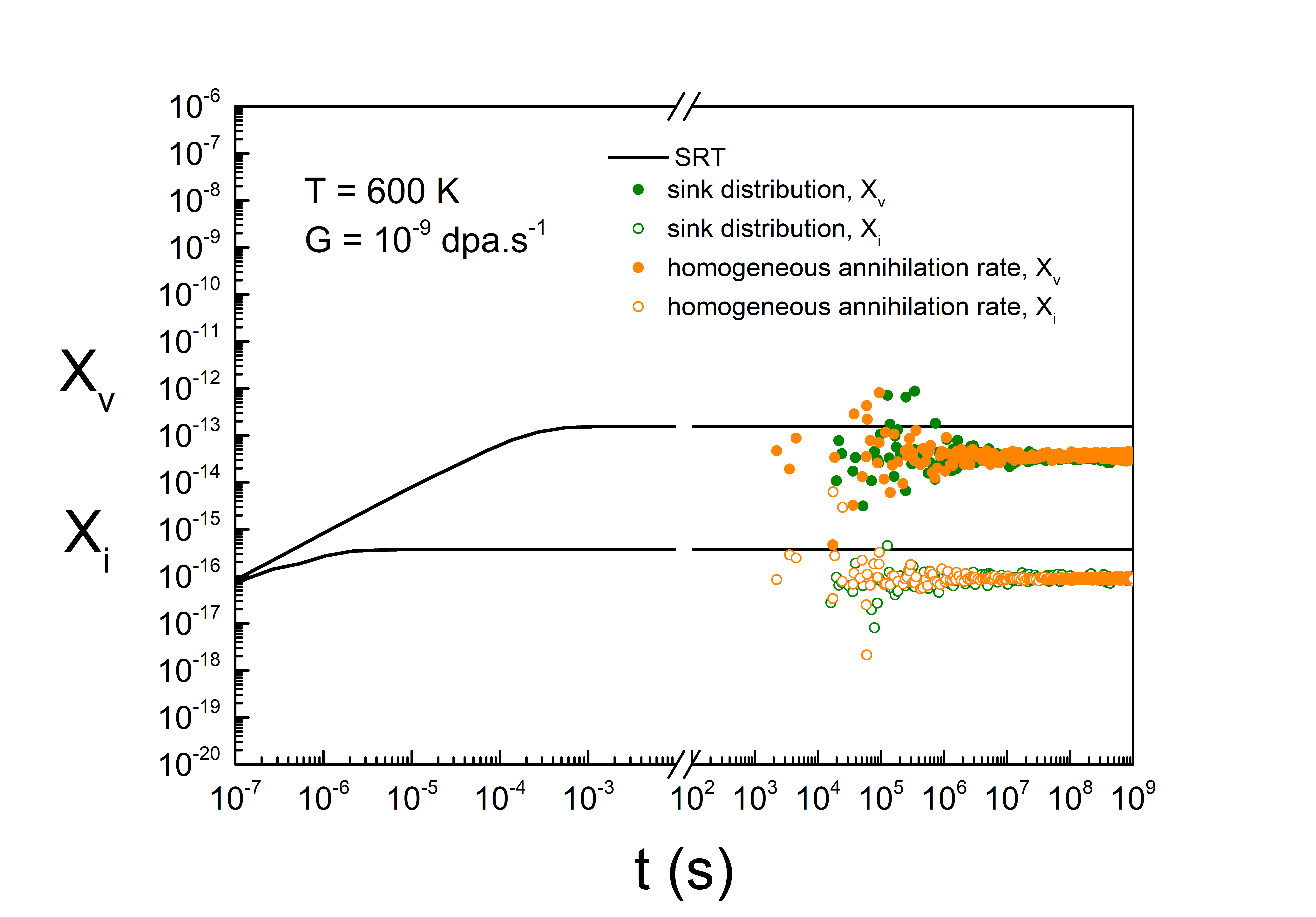}
\caption{Evolution of point defect concentrations at $T=600$~K and $G=10^{-9} \rm dpa.s^{-1}$, $k^2_ i=k^2_ v=5\times 10^{10}\rm cm^{-2}$. AKMC simulations with $R_p=20\lambda$ and $N_s=2\times64^3$, green dots: AKMC with a real distribution of sinks, orange dots: AKMC with an average elimination rate.}
\label{fig2}
\end{figure}

Examples for the evolution of $X_v(t)$ and $X_i(t)$, for different irradiation conditions, are shown in Figs.~\ref{fig1} and \ref{fig2}. The results obtained using both models of point defect elimination at sinks, give similar results. In both cases, measurements initially show large fluctuations, because for short times the averaged concentrations must be measured on a limited number of MCS (each dot corresponds to a value of $X_v$ or $X_i$ averaged on $10^5$ MCS at the beginning of the simulation, and on $10^8$ MCS at the end). The fluctuations become negligible at longer times. In general, transient regimes are difficult to model with reasonable system sizes (in a system containing initially no point defect, the minimum accessible time is $t_{min}=1/(N_sG)$, i.e. $\sim ~2 \times 10^{-3}$~s for Fig.~\ref{fig1} and $\sim ~2 \times 10^{3}$~s for Fig.~\ref{fig2}.

At high dose rates / low temperatures (Fig.~\ref{fig1}), one may nevertheless observe the end of the transient regime : the decrease on $X_i(t)$ when the elimination of interstitials at sinks becomes effective, driving an increase of $X_v(t)$ due to fewer recombination. The evolution of $X_v(t)$ and $X_i(t)$ measured in the AKMC simulation is in good agreement with the predictions of SRT (obtained by numerical integration of Eqs.~\ref{SRT2}).

For low dose rates / high temperatures, the point defects concentrations are much smaller and have their steady-state values as soon as they can be measured. The AKMC values are significantly smaller than the SRT prediction : this discrepancy is a finite-size effect which will be discussed in more detail in what follows.

\subsection{Steady-state point defect concentrations without point defect sinks}

\begin{figure}[htbp]
\centering
\includegraphics[width=1.0\columnwidth]{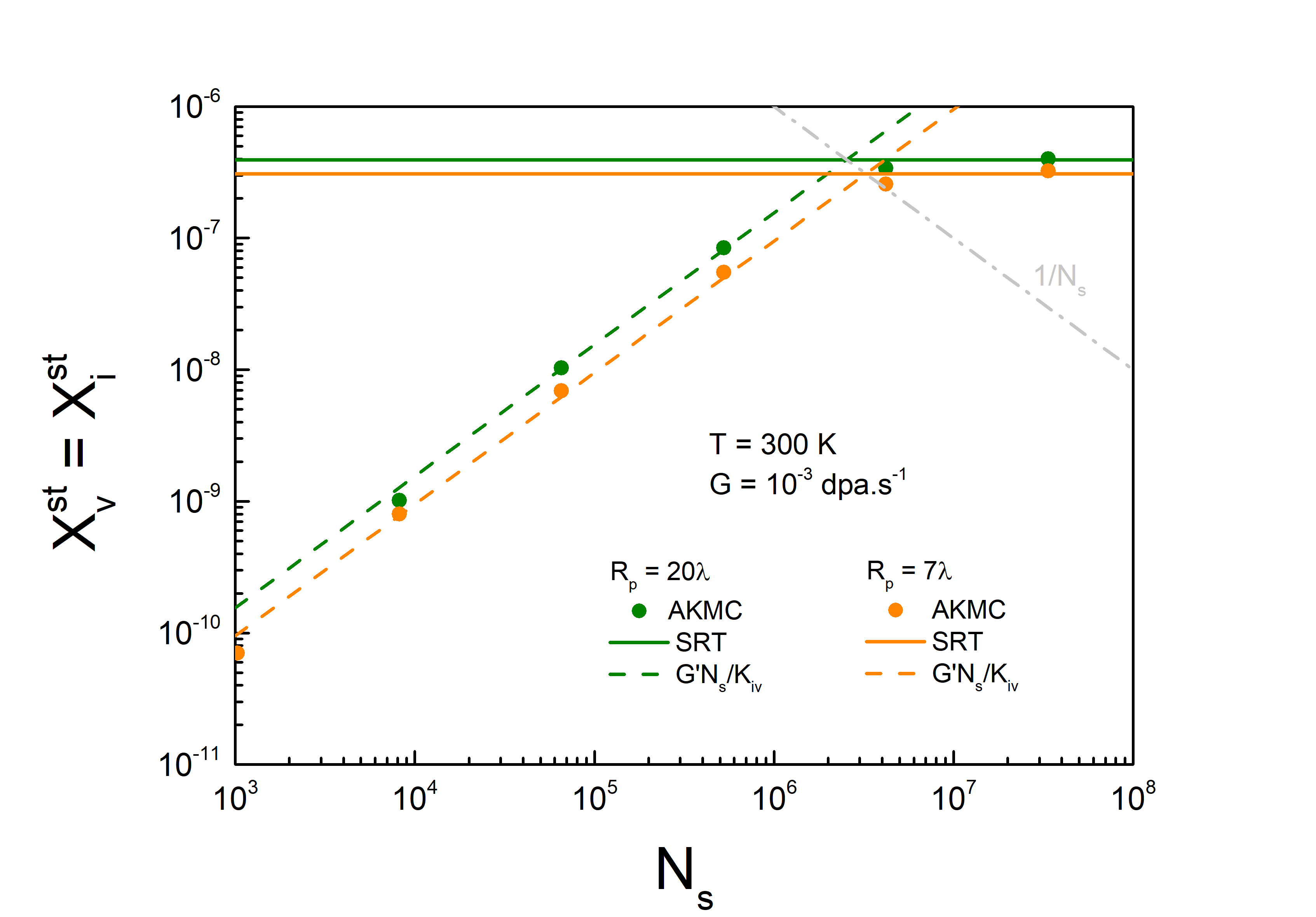}
\caption{Steady-state point defect concentrations in pure iron at 300K, $G=10^{-3} \rm dpa.s^{-1}$ and no sinks, as a function of the system size $N_s$ (number of bcc sites) as given by AKMC simulations and SRT. The figure also shows the effect of the correlated recombinations for $R_p = 20\lambda$ and $R_p=7 \lambda$.}
\label{fig3}
\end{figure}

Without point defect sinks ($K_v=K_i=0$), the steady-state solution of SRT is $X_v^{st}=X_i^{st}=\sqrt{G'/K_{iv}}$. The steady-state values of $X_v^{st}$ and $X_i^{st}$ measured in AKMC simulations with different size samples $N_s$, at $T=300$~K and $G=10^{-3} ~\rm dpa.s^{-1}$, are shown in Fig.~\ref{fig3}. For large systems, the AKMC results are in good agreement with the prediction of SRT, and independent of  $N_s$. As demonstrated in the literature \cite{Schroeder75}, the effect of close pair recombination results in a decrease in the effective production rate $G'=\xi G$, with $\xi=1-R_{rec}/R_p\sim 0.827$ and 0.505 for $R_p = 20$ and $7\lambda$, respectively. However, below a given size, AKMC simulations give $X_v^{st}$ and $X_i^{st}$ values which depend on the system size ($X_v^{st}=X_i^{st} \propto N_s$) and are significantly smaller than the SRT results.

\begin{figure}[htbp]
\centering
\includegraphics[width=1.0\columnwidth]{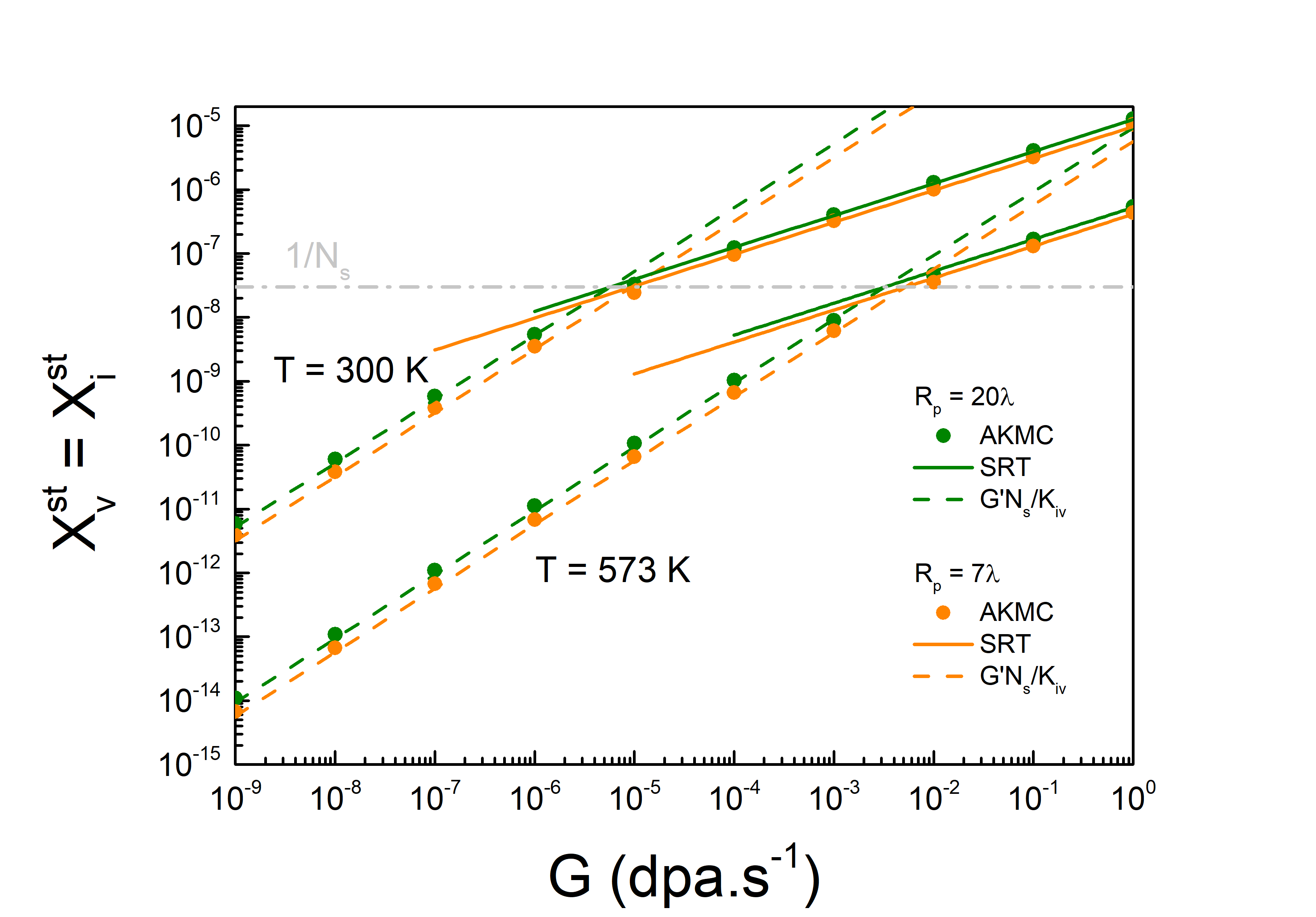}
\caption{Evolution of the steady-state concentration of point defects in pure iron at 300 and 573K, as a function of the production rate, in a system of $N_s = 2\times256^3$ bcc sites without sinks. Effect of the correlated recombination for  $R_p = 20\lambda$ and $R_p=7 \lambda$. Comparison between AKMC and the standard Rate Theory}
\label{fig4}
\end{figure}

We observe the same kind of discrepancy at various temperatures and production rates in AKMC simulations performed with a simulation box of $N_s = 2\times256^3$ bcc sites (Fig.~\ref{fig4}). At high dose rate $G$, the AKMC results are very close to the SRT results. At low $G$, the concentrations measured in the AKMC simulations are smaller than $\sqrt{G'/K_{iv}}$. One can see that $X_v^{st}$ and $X_i^{st}$ are then proportional to $G$ (or $G'$). For $G=10^{-6}~\rm dpa.s^{-1}$, the discrepancy reaches a factor $\sim 5$ at 300~K and $\sim 100$ at 573~K.

The origin of the discrepancy is easy to understand. When the spatio-temporal average values $X_v^{st}$ and $X_i^{st}$ drop below $1/N_s$, there is either no point defect in the simulation box, or one vacancy and one interstitial at the same time. SRT neglects the fact that these two defects are created together and disappear together. This correlation in time produces vacancy and interstitial pairs correlated in space, due to the finite size of the simulation box. We obtain an approximate solution for the concentration of these pairs by considering that in such conditions, a vacancy with the concentration $X_v$ may recombine with an interstitial with a concentration $1/N_s$. This leads to the steady state solution $X_v^{st}=X_i^{st}=G' N_s/K_{iv}$, in good agreement with the AKMC results (Fig.~\ref{fig3} and \ref{fig4}). A full description of these correlations will be given in section 4. 


\subsection{Steady-state point defect concentrations with point defect sinks}

\begin{figure}[htbp]
\centering
\includegraphics[width=1.0\columnwidth]{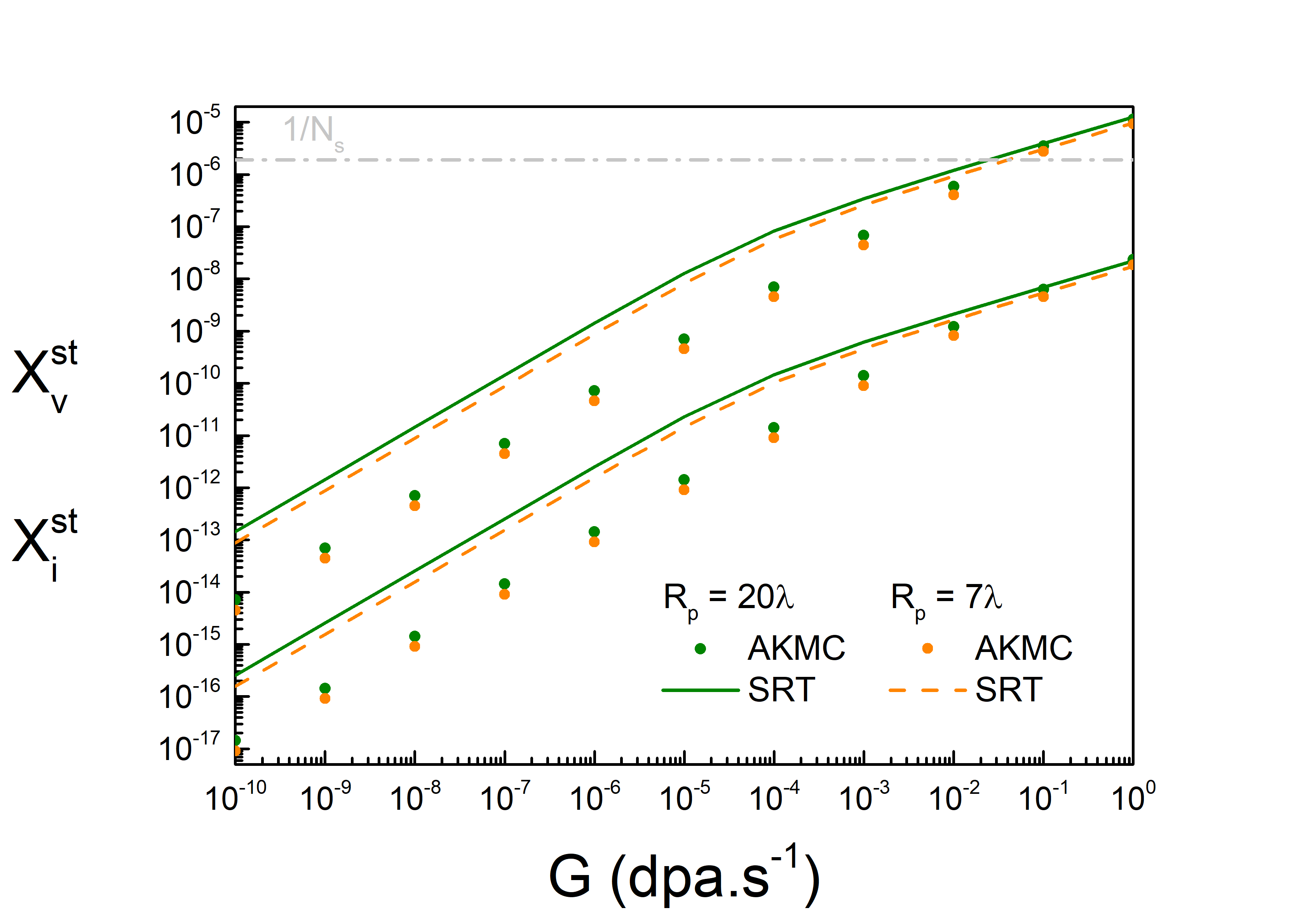}
\caption{Evolution of the steady-state concentration of point defects in pure iron at 573K as a function of the production rate in a system of $N_s = 2\times64^3$ bcc sites with a sink strength of $k_i^2=10^{10} \rm cm^{-2}$. Effect of the correlated recombination: $R_p = 20\lambda$ and $R_p=7 \lambda$. Comparison between AKMC and the SRT.}
\label{fig5}
\end{figure}

An example of evolution of the steady-state concentration of point defects in the presence of sinks (sink strength $k_i^2=10^{10} \rm cm^{-2}$) is given in Fig.~\ref{fig5}. AKMC simulations have been performed in a simulation box with $N_s = 2\times64^3$ bcc sites. At high dose rates ($>10^{-2}~\rm dpa.s^{-1}$), the concentrations measured in the simulations are in good agreement with the predictions of SRT (Eqs. \ref{SSC_V} and \ref{SSC_I}). Interstitials diffuse and reach the sinks more rapidly than vacancies, which leads to $X_i^{st}=X_v^{st} D_v/D_i \ll X_v^{st}$. As in the previous cases, close pair recombination only affects $X_i$ and $X_v$ by slightly reducing $G'=\xi G$ (always with $\xi \sim 0.827$ and 0.505 for $R_p = 20$ and $7\lambda$, respectively).

At lower dose rates the SRT overestimates the point defect concentrations. The discrepancy increases as $G$ decreases and reaches a constant values of $\sim 20$ below approximately $10^{-5}~\rm dpa.s^{-1}$. 

\section{Correlated Pair Theory} \label{toto}

The SRT model neglects any correlations in space and time that may arise due to physical correlations or  finite size effects. We introduce  a correlated pair theory (CPT) accounting for the correlations in time between defects of a Frenkel pair and finite size effects on the recombination reaction probability. 


In simulation boxes of reduced size, there is often a single Frenkel pair in the system. It is thus essential to account for the fact that most of the time an interstitial can only recombine with a vacancy simultaneously present in the simulation box. In order to account for these correlated events,  
we introduce two categories of point defect populations: (1) Interstitial-Vacancy correlated pairs $X_{iv}$ in which $i$ and $v$ are simultaneously present; and (2) mono-interstitial $X_i^{mo}$ and mono-vacancy $X_v^{mo}$ populations which are neither correlated in time or space
\begin{align}
\label{Total1}
X_v=&X_{v,mo}+X_{iv} \\  \nonumber
X_i=&X_{i,mo}+X_{iv}.
\end{align}
Note that the concentration of pairs is fixed by the smallest total concentration of point defects, between vacancies and self-interstitials. In the present case, the interstitial concentration is the smallest one. Thus, when the mobility of interstitial defects is significantly higher than that of vacancies we can approximate $X_{iv}=X_i$ and $X_{i,mo}=0$. 

Also notice that we do not explicitly account for the correlations in space of a close pair because these correlations are already taken into account within the definition of the effective production rate $G'$. As explained in the previous section, the probability of a mutual recombination between the point defects of a time-correlated pair is equal to the probability $X_{iv}$ of having one of the component of the pair on a given site multiplied by the probability of finding the other component of the pair on an adjacent site, which is equal to $1/N_s$. 
 Hence, we write for the rate equation of the time correlated pairs $X_{iv}$
\begin{align}
\label{Xcort}
\frac{d X_{iv}}{dt}=&+G' -K_{iv} X_{iv} \frac{1}{N_s} -K_{iv} (X_{iv})^2 \\ \nonumber
&-K_{iv} X_{iv}(X_{i,mo}+X_{v,mo}) \\ \nonumber
&-(K_i+K_v)X_{iv}.   
\end{align}
The third term in the RHS corresponds to recombination reactions between point defects belonging to two different pairs. We write the recombination rate after the standard Waite's formula because these distinct pairs are not correlated in time. Note that a recombination of two pairs leaves one vacancy and one interstitial correlated in time, forming then a new correlated pair. Therefore, after the reaction, only one correlated pair is removed, leading to $-K_{iv} X_{iv}^2$. The new pair is correlated in time, without being necessarily a close pair. Pairs can also be removed, by recombining with a mono-vacancy or mono-interstitial leading to the recombination term $-K_{iv} X_{iv}(X_i^{mo}+X_v^{mo})$. Eventually, a pair may be eliminated because the vacancy or interstitial forming the correlated pair is annihilated at sinks. This is represented by the last term in the RHS.

In the same way, we write the rate equations of the mono-species
\begin{align}
\label{XV}
\frac{d X_{v,mo}}{dt} &=-K_{iv} X_{v,mo}  X_{i,mo}  + K_i X_{iv} \nonumber \\
&- K_v X_{v,mo},
\end{align}
and
\begin{align}
\label{XI}
\frac{d X_{i,mo}}{dt} &=-K_{iv} X_{v,mo}  X_{i,mo} + K_v X_{iv} \nonumber \\
&- K_i X_{i,mo}.
\end{align}
A mono-vacancy (mono-interstitial) is removed when recombined with a mono-interstitial (mono-vacancy) or when eliminated at sinks. A mono-vacancy (mono-interstitial) is formed after the elimination at sinks of an interstitial (vacancy) belonging to a correlated pair, leading to a positive reaction term, $K_i X_{iv}$ ($K_v X_{iv}$). Note that the recombination of a correlated pair with a mono-species leads to a zero balance for the mono-species:
for instance, a mono-vacancy recombining with an interstitial of a correlated pair leaves a mono-vacancy.
The sums of Eqs.~\ref{Xcort} and \ref{XV} for the vacancy; and Eqs.~\ref{Xcort} and \ref{XI} for the interstitial, lead to the system of equations
\begin{align}
\label{dXvtot}
\frac{d X_{v}}{dt}=&+G' -K_{iv} X_{iv} \frac{1}{N_s} -K_{iv} X_i X_v -K_v X_v,
\end{align}
and
\begin{align}
\label{dXitot}
\frac{d X_{i}}{dt}=&+G' -K_{iv} X_{iv} \frac{1}{N_s} -K_{iv} X_i X_v -K_i X_i.
\end{align}
These equations are equivalent to the SRT ones when there is no size effects ($N_s$ tends to infinity).

In order to derive analytical expressions of the stationary point defect populations, we distinguish two different kinetic regimes.
First, we ignore the elimination of point defects at sinks and second, we assume
the elimination at sinks is the dominant mechanism.

\subsection{Correlated pairs with no elimination at point-defect sinks} \label{nosink}
If we ignore the point defect sinks ($K_v=K_i=0$), the stationary condition applied to Eqs.~\ref{XV} and \ref{XI} implies
\begin{equation}
X_{v,mo}^{st}=X_{i,mo}^{st}=0.
\label{monomers}
\end{equation}
From Eqs.~\ref{Xcort} and ~\ref{monomers}, we obtain the steady state pair concentration $X_{iv}^{st}$ as a solution of the second-order polynomial equation
\begin{equation}
G'-X_{iv}^{st} \frac{1}{N_s} -{X_{iv}^{st}}^2 = 0.
\label{Xiv-t-ST-f}
\end{equation}
The single physical solution writes
\begin{align}
\label{Xivrec}
X_{iv}^{st}= 
\sqrt{\frac{G'}{K_{iv}}+\left(\frac{1}{2N_s}\right)^2}-\frac{1}{2N_s}, 
\end{align}

The total vacancy and self-interstitial concentrations are equal to the concentration of correlated pairs.
As expected, Eq.~\ref{Xivrec} tends to the SRT solution
when there is no finite size effects. 
The finite size effects are negligible as long as $1/N_s \ll \sqrt{G'/K_{iv}}$,  which after Eqs.~\ref{monomers} and \ref{Xivrec} is similar to the condition $1/N_s \ll X_v^{st}= X_i^{st}$. Therefore, the threshold value of the radiation dose rate $G$, at which there is no more finite size effects depends on temperature through the variation of $K_{iv}$ with temperature.
As presented in Fig.~\ref{fig6}, the CPT results yield an increase with temperature of the radiation dose rate threshold value in excellent agreement with the AKMC simulations. Furthermore, as predicted by the CPT, deviations between SRT and the CPT-AKMC results start when the concentration of point defects is below $1/N_s$.

\subsection{Correlated pairs with annihilation at sinks} \label{sink}

When point defects eliminate at sinks, the population of correlated pairs and mono-species coexist.
A stationary condition applied to Eqs.~\ref{XV} and \ref{XI}, leads 
to the standard relationship between the total
concentrations of $i$ and $v$
\begin{equation}
\label{steady}
K_i (X_{iv}^{st}+X_{i,mo}^{st})=K_v (X_{iv}^{st}+X_{v,mo}^{st}).
\end{equation}
The concentration of pairs $X_{iv}$ is either equal to the total concentration of vacancies $X_v$ or interstitials $X_i$, depending
on their relative amplitudes. In the present case, the diffusion coefficient of interstitials is much higher than the one of vacancies. 
Therefore, we have $K_i \gg K_v$ and $X_v \gg X_i$, leading to $X_{iv}^{st}\approx X_i^{st}$ and $X_{i,mo}^{st}\approx 0$. By solving Eqs.~\ref{dXvtot}, ~\ref{dXitot}, and ~\ref{steady},
we obtain the steady state total concentrations $X_i^{st}$ and $X_v^{st}$ 
\begin{align}
X_v^{st}&=-\frac{K_i}{2K_{iv}}\left(1+\frac{K_{iv}}{K_i N_s}\right) \nonumber \\
&+\left[\left(\frac{K_i}{2K_{iv}}\right)^2 \left(1+\frac{K_{iv}}{K_i N_s}\right)^2+\frac{G'K_i}{K_{iv}K_v}\right]^{1/2},
\label{SSC_VCPT}
\end{align}
and
\begin{align}
X_i^{st}&=-\frac{K_v}{2K_{iv}} \left(1+\frac{K_{iv}}{K_i N_s}\right) \nonumber \\
&+\left[\left(\frac{K_v}{2K_{iv}}\right)^2 \left(1+\frac{K_{iv}}{K_i N_s}\right)^2+\frac{G'K_v}{K_{iv}K_i}\right]^{1/2}.
\label{SSC_ICPT}
\end{align}
As expected, in the absence of size effects, Eqs.~\ref{SSC_VCPT} and \ref{SSC_ICPT} are similar to Eqs.~\ref{SSC_V} and \ref{SSC_I}, and the steady state concentrations are the SRT ones. Size effects systematically reduce the
total concentration of point defects. 
We deduce the concentration of  mono-vacancies from a steady state condition applied to Eq.~\ref{XV}
\begin{equation}
 X_{v,mo}^{st}=\frac{K_i}{K_v} X_{i}^{st}.
\label{Xvmo}
\end{equation}
Note that the partition between monomers and pairs does not depend on the size of the system, but only on the ratio $K_v/K_i$. Even in a large system with no finite size effects, there is a population of mono-species and time correlated pairs, whereas the total steady state concentrations of vacancies and interstitials are the ones predicted by SRT.

We deduce from the steady state point defect concentrations, 
the ratio of the total point defect recombination rate divided by the elimination rate of both vacancies and self-interstitials
\begin{equation}
\frac{n_{rec}^e}{n_{sink}^e} = \frac{K_{iv}}{4 K_i K_v} (K_v X_v + K_i X_i)+\frac{K_{iv}}{2 N_s (K_i+K_v)}.
\label{ratiocor2}
\end{equation}
This ratio is sensitive to finite size effects. In cases where size effects are negligible, this ratio corresponds to the one given by SRT. The smaller the number of sites $N_s$, the higher the ratio. Therefore, small systems promote the recombination reactions with respect to the annihilation reactions at sinks. Interestingly, finite size effects on this ratio depend on the radiation dose rate $G$ through the variation of point defect concentration with $G$.
 In Fig.~\ref{fig7}, we observe an excellent agreement between the AKMC simulations and the CPT results, both for the concentration of point defects and the recombination/annihilation ratio.

\section{Modified AKMC simulations} \label{Modified AKMC}

As it was mentioned above, the AKMC naturally accounts for temporal and spatial correlations. This implies that the results, in terms of defect concentrations, from simulations using small boxes will deviate from the SRT. Due to such correlations, the latter underestimates the recombination probabilities, which leads to higher concentrations compared to AKMC. In that respect, the AKMC values of point defect concentrations are the correct ones. But the SRT values are inaccurate only because by using small system sizes one introduces correlations that would have a negligible effect in a real (i.e. large) system. 

In the following we develop expressions for the probability of recombination (or equivalently rejection probabilities) of a given recombination event, that can be used in the AKMC to give the concentration evolution that would be obtained in large samples, i.e. equivalent to the SRT.

\subsection{Recombination probability without sinks}

\begin{figure}[htbp]
\centering
\includegraphics[width=1.0\columnwidth]{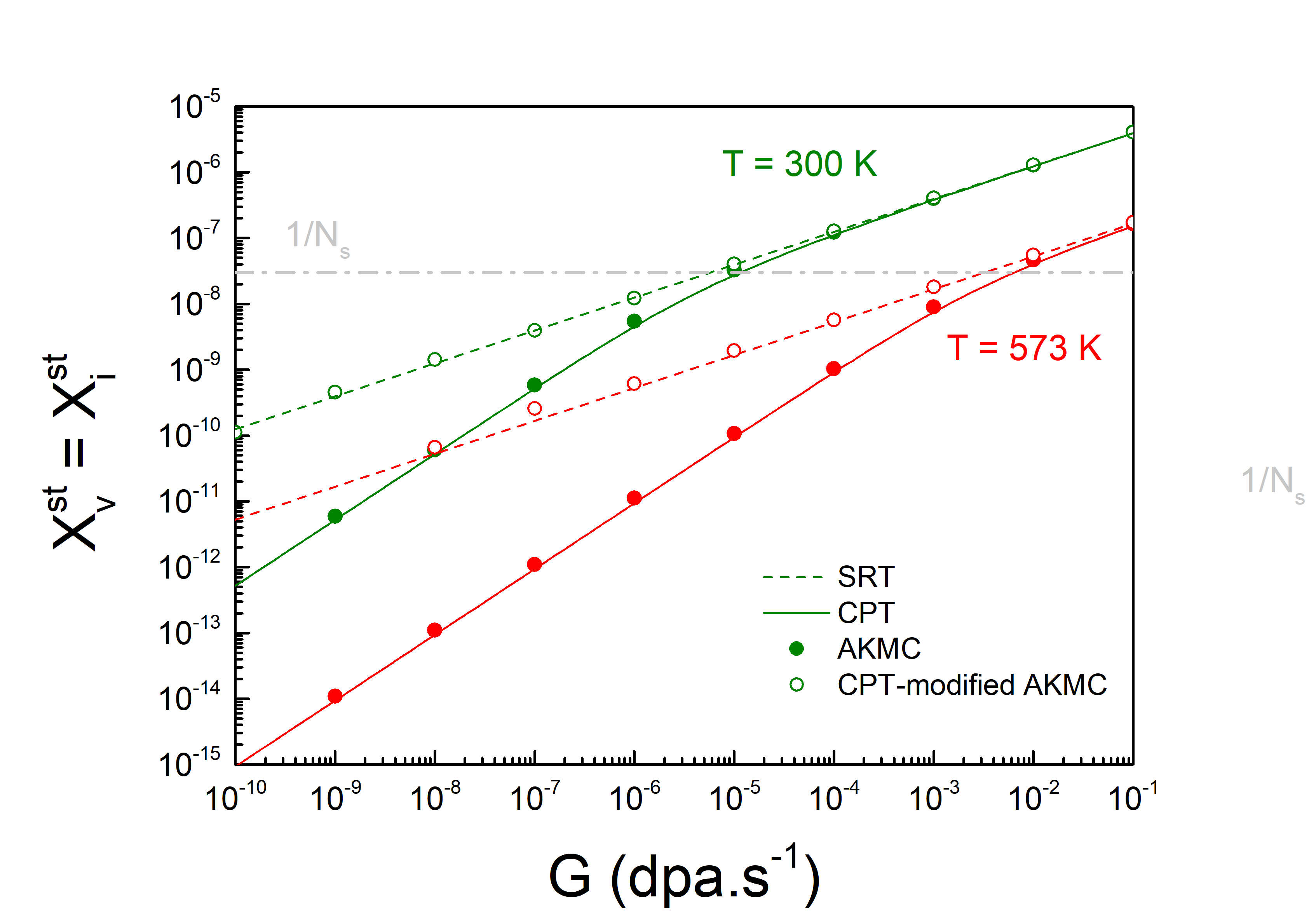}
\caption{Steady-state point defect concentration in pure iron at 300 and 573K as a function of the production rate in a system with $N_s = 2\times256^3$ bcc sites without sinks as given by Standard Rate Theory, Correlated Pair Theory, and AKMC simulations with and without CPT corrections ($R_p= 20\lambda$).}
\label{fig6}
\end{figure}

In case there are no sinks in the system, we can define the recombination probability to obtain the correct defect concentrations as
\begin{equation}\label{eq:rejProb}
P(Rec)=\frac{X_{v}^{st,CPT}}{X_{v}^{st,SRT}}=\frac{X_{i}^{st,CPT}}{X_{i}^{st,SRT}}
\end{equation}
where $X_{v}^{st,CPT}=X_{i}^{st,CPT}$ are given in Eq.~\ref{Total1} and $X_{v}^{st,SRT}=X_{i}^{st,SRT}$ in Eq.~\ref{eq:RTNoS}. Substituting these expressions into Eq.~\ref{eq:rejProb} we obtain
\begin{align}\label{eq:rejProbFin}
P(Rec| R \leq R_{rec}) &=
    \frac{\sqrt{\frac{G'}{K_{iv}}+\frac{1}{4N_s^2}}-\frac{1}{2N_s}}{\sqrt{\frac{G'}{K_{iv}}}} \nonumber \\
    &= \frac{1}{2N_s}\left [ \frac{\sqrt{\frac{4G'N_s^2+K_{iv}}{K_{iv}}}-1}{\sqrt{\frac{G'}{K_{iv}}}} \right ]
\end{align}
which is the recombination probability given that defects are inside the recombination distance, i.e., the conditional probability that, provided that defects are within the recombination distance, the recombination is actually performed. To sample this distribution, a random number in the range $[0:1)$ is drawn from a uniform distribution. If it is smaller than $P(Rec| R \leq R_{rec})$ the recombination takes place, otherwise the defect that last moved is placed at a distance $R_p$ following the methodology described in Section~\ref{Monte Carlo Simulations and finite size effects}. Results from this approach are shown in Fig.~\ref{fig6} (using the same conditions as for Fig.~\ref{fig4}), where we see that the open circles match the SRT results (dotted line). 

\subsection{Recombination probability with sinks}

In case sinks are present, the methodology is the same but the equations for the concentration of defects change. We apply the defined probability to both defect recombination and annihilation at sinks. The expressions for the recombination probability depend in this case on the specific defect that jumped the last before checking for recombination.
\begin{equation}
    P_v(Rec)=\frac{X_{v}^{st,CPT}}{X_{v}^{st,SRT}},\quad P_i(Rec) = \frac{X_{i}^{st,CPT}}{X_{i}^{st,SRT}}
\end{equation}
where we substitute Eq.~\ref{SSC_V} for $X_{v}^{st,SRT}$ and Eq.~\ref{SSC_VCPT} for $X_{v}^{st,CPT}$ for vacancies and Eq.~\ref{SSC_I} for $X_{i}^{st,SRT}$ and Eq.~\ref{SSC_ICPT} for $X_{i}^{st,CPT}$ for self-interstitials. 


\begin{figure}[htbp]
\centering
\includegraphics[width=1.0\columnwidth]{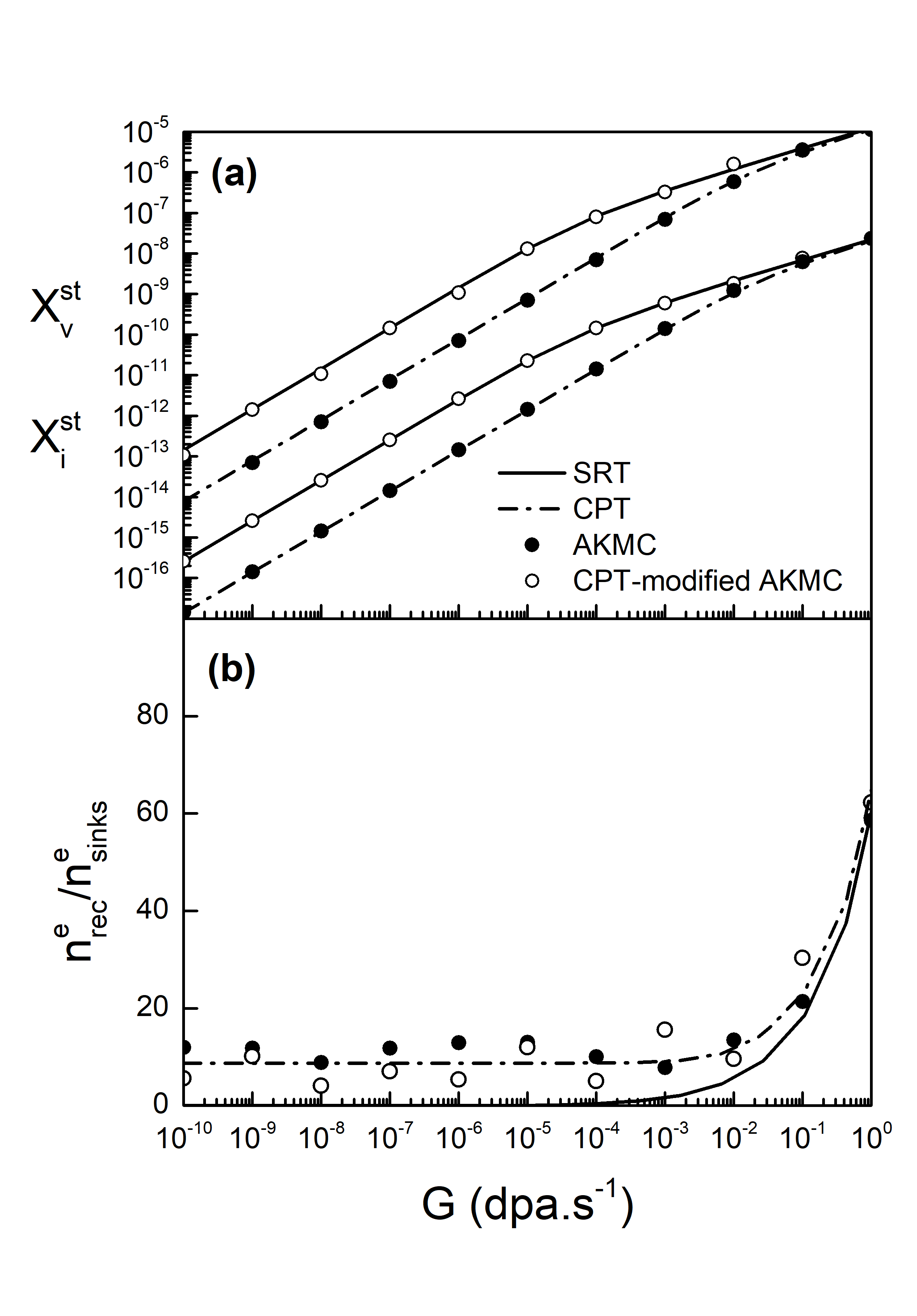}
\caption{Evolution of (a) the steady-state concentration of point defects and (b) 
the ratio of recombination to elimination at sinks, in pure iron at 573K, as a function of the production rate, in a system of $N_s = 2\times64^3$ bcc sites with a sink strength of $k_i^2=10^{10} ~\rm cm^{-2}$ ($R_p = 20\lambda$).  Standard Rate Theory, Correlated Pair Theory, AKMC simulations with and without CPT corrections.}
\label{fig7}
\end{figure}

Figure ~\ref{fig7}(a) shows AKMC results, with the rejection probability matching the SRT predictions. Figure ~\ref{fig7}(a) presents the ratio between the number of recombinations and annihilations at sinks. It is worth noting that the proposed recombination probability does not result in the same ratio as predicted by SRT, as it retains the recombinations to annihilations ratio as given by the standard AKMC. It is also important to mention that the quantity that controls the microstructure evolution is the defect concentration, and the gradients of such. Hence, we conjecture that the main physics will be captured by the proposed correction.

\section{Discussion and Conclusions} \label{Conclusions}

It is important to emphasize that the relevance of finite size effects, as shown by the CPT, strongly depends on the irradiation conditions, the point defect properties, and (obviously) the system size. We have chosen the parameters  of  Fig.~\ref{fig7} in such a way to maximize the effects of correlated recombination, by using a relatively small box ($N_s = 2\times64^3$) and introducing a low sink density $k_i^2=10^{10} ~\rm cm^{-2}$). In this extreme case, the AKMC and SRT concentrations differ by a factor of $\sim 20$.
Keeping the same size $N_s = 2\times64^3$ and increasing the sink strength lead to a significant decrease of the SRT-AKMC discrepancy (because less point defects recombine): only a factor of $\sim 3$ for $k_i^2=10^{11}~ \rm cm^{-2}$, and  $\sim 17\%$ for $k_i^2=10^{12} ~\rm cm^{-2}$.
Or maintaining the same sink density ($k_i^2=10^{10} ~\rm cm^{-2}$), one obtains a better SRT-AKMC agreement by increasing the size of the system: the discrepancy decreases to a factor of $\sim 3$ for $N_s = 2\times128^3$ , and to $0.27$ for $N_s = 2\times256^3$.

In case we introduce point-like sinks explicitly and their corresponding atomic fraction is smaller than the inverse of the size of the simulation box, we expect finite size effects on the elimination rate of point defects at sinks as well. We can extend the CPT theory to account for the correlations in time between defects of the Frenkel pairs and point-like sinks. To generalize our approach to systems featuring large heterogeneities of point defect and/or solute concentration fields, we could choose a reference bulk region in the system, treat the surrounding as a continuum from which we extract an effective point defect sink strength. Then, we could use the CPT theory to calibrate the point defect recombination rate of point defects in the reference bulk region. Another perspective will be to rely on the CPT theory  to investigate the irradiation and microstructure conditions for which finite size effects can be expected.

Also important to note is the fact that the correction to the AKMC point defect elimination relies on the steady-state defect concentrations. Hence, this approach does not guarantee that the transients are accurate, only that the steady-state concentrations are in agreement with the SRT. We do not expect the transients to be significantly distinct, but its accurate quantification is still an open problem.

A possible strategy to deal with these finite size effects, is to first use the CPT to see if the correlated recombination affects the results. If they do, try to increase the size of the simulation box (which will lead to higher CPU times), and if the required CPU cost is overwhelming, use the CPT-modified AKMC method. 


In this paper, we have developed an extension of the SRT formalism for systems under irradiation to consider the interplay between the spatio-temporal correlations between point defects and finite size effects. We demonstrate that these correlations can lead to significant discrepancies in the concentration of defects when the size of the system sizes is small, as it is usually the case in kinetic Monte Carlo (KMC) simulations. This novel framework, the so-called Correlated Pair Theory (CPT), is based on the explicit introduction of the concentration of correlated Frenkel pairs in the rate theory expressions (see Eqs.~\ref{Xcort}, \ref{XV} and \ref{XI}). We show how the CPT theory reproduces remarkably well the results from AKMC simulations in systems of reduced size, hence capturing the main effects of the spatio-temporal correlations. CPT holds for both regimes, where defect recombination dominates or defect annihilation at sinks dominates.

We have also developed recombination probability expressions to be used within an AKMC methodology to recover the SRT results. These expressions are based on the \textit{a priori} knowledge of the steady-state concentrations given by the SRT formalism and the CPT. We show that the proposed modification of the defect annihilation within the AKMC leads to point defect concentrations similar to those given by the SRT approach. However, it does not recover the SRT ratio between defect recombination and annihilation at sinks, since it retains the ratio between recombination and annihilation given by the standard AKMC. We argue that the total concentration is the important quantity to capture as it dominates the microstructure evolution.

\section{Acknowledgments} \label{Acknowledgments}
EM gratefully acknowledges support from the U.S. DOE, Office of Science, Office of Fusion Energy Sciences, and Office of Advanced Scientific Computing Research through the Scientific Discovery through Advanced Computing (SciDAC) project on Plasma-Surface Interactions (award no. DE-SC0008875). The research leading to these results has been carried out in  the frame of EERA Joint Program for Nuclear Materials and is partly funded by the European Commission HORIZON 2020 Framework Programme under grant agreement No. 755269.

\section{Data Availability}
All data used in the simulations performed in this paper is available upon request.

\bibliographystyle{elsarticle-num}
\bibliography{bib}

\end{document}